\documentclass[journal=jctc,manuscript=article]{achemso}
\usepackage[version=3]{mhchem} 
\usepackage{xcolor}

\author{Maxim Sukharev$^{\ddag}$}
\affiliation{College of Integrative Sciences and Arts, Arizona State University, Mesa, Arizona 85212, USA}
\affiliation{Department of Physics, Arizona State University, Tempe, Arizona 85287, USA}
\email{maxim.sukharev@asu.edu}
\author{Joseph E. Subotnik}
\affiliation{Department of Chemistry, Princeton University, Princeton, New Jersey 08540, USA}
\author{Abraham Nitzan}
\affiliation{Department of Chemistry, University of Pennsylvania, Philadelphia, Pennsylvania 19104, USA}

\title[An \textsf{achemso} demo]
{Unveiling the Dance of Molecules: Ro-Vibrational Dynamics of Molecules under Intense Illumination at Complex Plasmonic Interfaces}

\begin{document}
	
	\begin{abstract}
		Understanding the quantum dynamics of strongly coupled molecule-cavity systems remains a significant challenge in molecular polaritonics. This work develops a comprehensive self-consistent model simulating electromagnetic interactions of diatomic molecules with quantum ro-vibrational degrees of freedom in resonant optical cavities. The approach employs an efficient numerical methodology to solve coupled Schr$\ddot{\text{o}}$dinger-Maxwell equations in real space-time, enabling three-dimensional simulations through a novel molecular mapping technique. The study investigates relaxation dynamics of an ensemble of molecules following intense resonant pump excitation in Fabry-Perot cavities and at three-dimensional plasmonic metasurfaces. The simulations reveal dramatically modified relaxation pathways inside cavities compared to free space, characterized by persistent molecular alignment arising from cavity-induced rotational pumping. They also indicate the presence of a previously unreported relaxation stabilization mechanism driven by dephasing of the collective molecular-cavity mode. Additionally, the study demonstrates that strong molecular coupling significantly modifies the circular dichroism spectra of chiral metasurfaces, suggesting new opportunities for controlling light-matter interactions in quantum optical systems.
	\end{abstract}
	
	\section{Introduction} \label{introduction}
	The ability to control quantum molecular processes through electromagnetic (EM) fields has been a longstanding goal in chemistry. Famous coherent control schemes\cite{Shapiro2003} are often applied to guide individual molecules through desired photochemical pathways via various elaborate approaches, including laser pulse shaping.\cite{Rice2000,Brixner2001} Such ideas showed initial promise,\cite{Brixner2003} but their practical implementation has faced significant challenges due to decoherence\cite{Branderhorst2008}, limited selectivity\cite{Rabitz2000}, and highly complex optimal fields that are often difficult to implement experimentally. This challenge is compounded by the fact that most chemical reactions are collective (intermolecular) in nature, and achieving control of a single molecule does not necessarily guarantee control over the entire ensemble.
	
	Polaritonic chemistry,\cite{Ebbesen2023} emerging at the intersection of quantum optics and physical chemistry, offers another approach and relies on altering the EM environment that, in turn, collectively shapes chemical dynamics. It has revolutionized our understanding of light-matter interactions by exploiting the strong coupling regime between molecular transitions and spatially confined EM modes.\cite{Nagarajan2021} In this regime, the rate of collective energy exchange between molecules and a cavity mode exceeds their respective damping rates, leading to the formation of hybrid light-matter states called polaritons.\cite{Torma2015} These hybrid states exhibit properties distinct from their uncoupled constituents, offering new pathways for controlling chemical reactivity and molecular dynamics.
	
	The experimental realizations of strong coupling have been predominantly achieved in Fabry-Perot optical microcavities,\cite{Wright2023} where molecular ensembles are interrogated between highly reflective mirrors.\cite{Simpkins2023} These cavities confine EM modes to small volumes, enhancing the coupling strength between corresponding electronic and/or vibrational molecular transitions and cavity photons. Early demonstrations focused on electronic strong coupling, leading to modifications in photochemical reaction rates and energy transfer processes.\cite{Hutchison2012} Subsequent investigations revealed that vibrational strong coupling could also be achieved, particularly in the infrared region, where molecular vibrations couple to cavity modes, potentially offering selective control over chemical reaction coordinates.\cite{Ahn2023}
	
	Despite significant experimental progress, theoretical description faces several fundamental challenges.\cite{Feist2018,Ribeiro2018,Hertzog2019,Fregoni2022} The primary obstacle is the exponential scaling of computational complexity with the number of molecules involved in collective strong coupling. Traditional quantum chemistry methods, designed for single-molecule calculations, become computationally intractable when applied to systems containing thousands or millions of strongly coupled molecules. Notwithstanding these colossal computational challenges, simulations of a large number of molecules have been shown to be feasible for simple Fabry-Perot geometries.\cite{Luk2017,Tichauer2021} The dynamics of nuclei in this approach, however, is treated classically, which may miss important effects as shown recently.\cite{SukharevJCP2023}
	
	Recently, plasmonic metasurfaces have emerged as a powerful alternative platform for molecular control,\cite{Cohn2021,Sufrin2024} offering unprecedented precision in engineering local EM environments. Modern nanofabrication techniques enable the realization of metasurfaces with exquisite spatial control over their optical properties,\cite{Kuznetsov2024} allowing for tailored light-matter interactions at the nanoscale. These engineered surfaces can generate highly confined EM fields with specific polarization states, field gradients, and chiral properties,\cite{Urban2019} providing unique opportunities for selective molecular excitation and control that surpass traditional coherent control approaches. A particular challenge in understanding molecule-metasurface interactions has been the computational treatment of many-molecule quantum dynamics coupled to complex EM environments. Traditional approaches scale poorly with the number of molecules driven by highly spatially varying EM fields, limiting their applicability to realistic systems. Our work addresses this limitation through an efficient molecular mapping technique\cite{Sukharev2023} that, combined with direct numerical integration of Maxwell's equations, enables simulations of large molecular ensembles while retaining a full quantum mechanical description of their rotational and vibrational degrees of freedom.
	
	The interaction between chiral EM modes and molecular quantum states is of special interest, as it opens new possibilities for enantioselective chemistry and spectroscopy.\cite{Biswas2024} There have been several intriguing reports on how chiral molecules influence polariton states.\cite{Guo2021,Chen2022,Riso2023,Baranov2023} Our approach captures the local handedness of EM modes and their interaction with molecular transitions, providing insights into how strong coupling modifies both molecular dynamics and the optical response of the metasurface, including circular dichroism spectra. By solving coupled Schr$\ddot{\text{o}}$dinger-Maxwell equations in real space and time, our method accounts for the full complexity of strong coupling between molecular ensembles and surface EM modes. This enables us to investigate previously inaccessible regimes of light-matter interaction, where collective effects and quantum coherences play crucial roles in determining system dynamics.
	
This paper is organized as follows. First, we provide details of the molecular model employed. Next, we consider molecules in a simple one-dimensional Fabry-Perot cavity and discuss the implications of vibrational and rotational degrees of freedom on strong coupling and electron relaxation. Finally, we present systematic studies of rotational and vibrational molecular dynamics at three-dimensional plasmonic interfaces.
	
	\section{Molecular model with quantized vibrations and rotations} \label{molecules}
\begin{figure}
		\centering
		\includegraphics[width=1.0\linewidth]{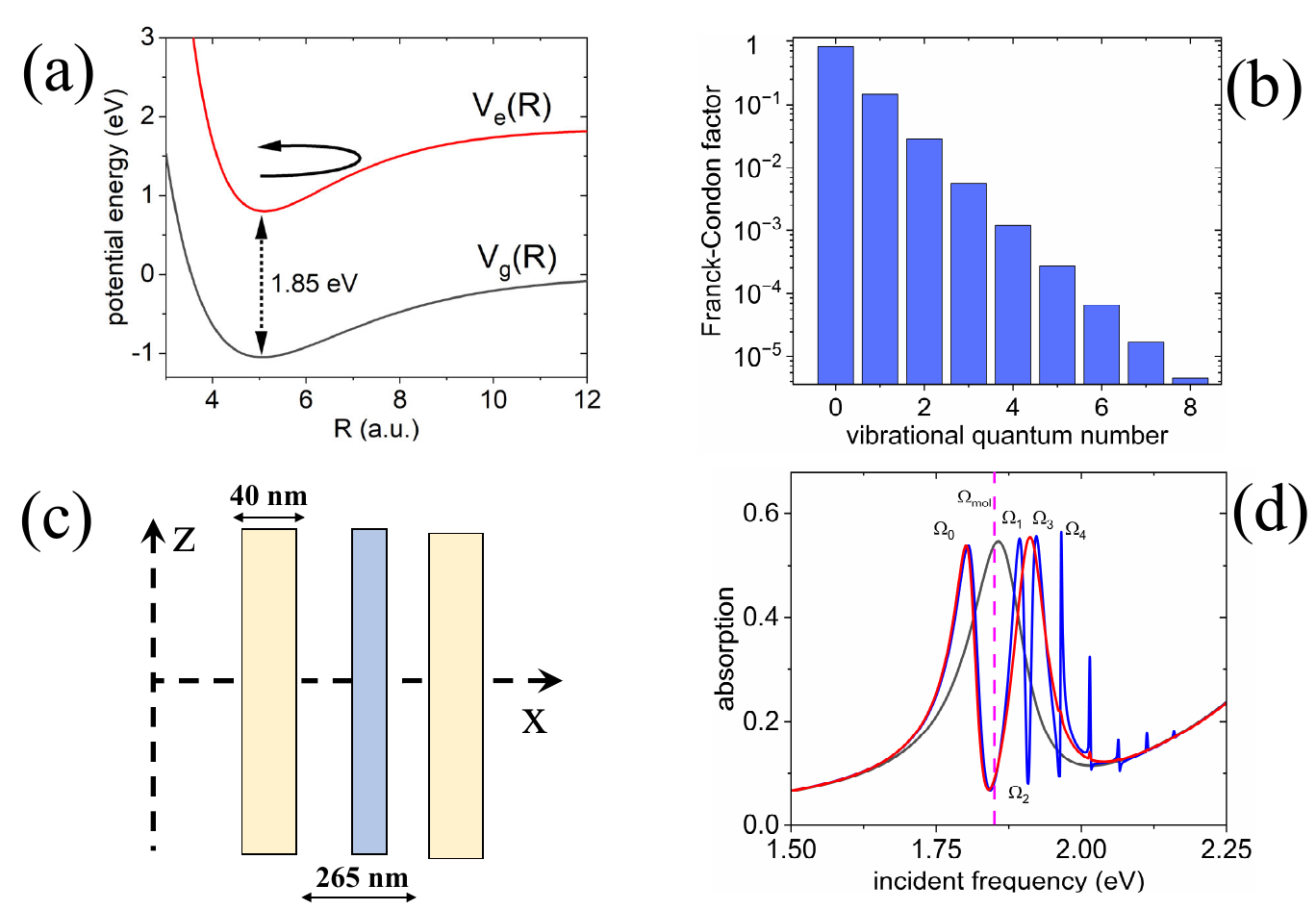}
		\caption{(a) Potential energy surfaces (PESs), with the $0-0$ transition energy at 1.85 eV. (b) Franck-Condon factors calculated by projecting the ground ro-vibrational state onto the vibrational states of the excited electronic manifold. (c) Schematic of the one-dimensional simulation setup: an 11 nm wide molecular slab is positioned at the center of an optical cavity formed by two thin Au slabs. The cavity dimensions are optimized to match its fundamental mode frequency with the molecular $0-0$ electronic transition. Probing and pumping fields are propagated along the $x$-axis with the incident field polarized along $z$. (d) Linear absorption spectra for three configurations: empty cavity (black line), cavity containing molecules with identical ground and excited PESs (no horizontal shift and identical Morse parameters; red line), and cavity containing molecules with PESs shown in panel (a) (blue line). The magenta dashed line indicates the $0-0$ transition frequency. Frequencies $\Omega_n$ mark the pumping frequencies used to simulate data shown in Figs. \ref{fig:2} and \ref{fig:3}. The molecular number density is $n_M=10^{26}$ m$^{-3}$.}
		\label{fig:1}
	\end{figure}

	We consider an ensemble of homonuclear diatomic molecules characterized by two potential energy surfaces (PESs), $V_g$ and $V_e$, corresponding to the ground and first excited electronic states, respectively. Both PESs support bound ro-vibrational states, as illustrated in Fig. \ref{fig:1}a. The potential energy surfaces are parameterized using a Morse function of the form
	\begin{eqnarray}
		\label{PESs}
		V_g(R)=D_g\left(1-\exp{\left(-a_g\left(R-R_g\right)\right)}\right)^2-D_g, \\
		V_e(R)=D_e\left(1-\exp{\left(-a_e\left(R-R_e\right)\right)}\right)^2+\Delta E,
	\end{eqnarray}
	where $R$ denotes the internuclear separation. The potential parameters used in the calculations reported below are $D_g=38.6$ meV, $a_g=0.458$ a.u., $R_g=5.06$ a.u., $D_e=D_g$, $a_e=0.587$ a.u., $R_e=5.26$ a.u., and $\Delta E=1.8114$ eV. The ground state potential corresponds to that of Li$_2$,\cite{Charron2006} while the excited state potential is slightly narrower, with parameters chosen to ensure the Franck-Condon distribution (shown in Fig. \ref{fig:1}b) is dominated by the $0-0$ transition.
	
	The total molecular wavefunction following the Born-Oppenheimer approximation is expressed as
	\begin{equation}
		\label{BO-appr}
		\Psi\left(\boldsymbol{r},\boldsymbol{R},t\right)=\Phi_g\left(\boldsymbol{r},R  \right)\frac{\chi_g\left( \boldsymbol{R},t\right)}{R}+\Phi_e\left(\boldsymbol{r},R  \right)\frac{\chi_e\left( \boldsymbol{R},t\right)}{R},
	\end{equation}
	where $\Phi_{g,e}$ represent the electronic wavefunctions for the ground and first excited electronic states, respectively. The coordinates $\boldsymbol{r}\equiv(r,\hat{r})$ and $\boldsymbol{R}\equiv(R,\hat{R})$ denote electronic and nuclear coordinates, including their angular dependence in the chosen coordinate system. For our homonuclear molecule, the electronic states possess specific symmetries: $^1\Sigma_g^+$ for the ground state $|g\rangle$ and $^1\Sigma_u^+$ for the first excited state $|e\rangle$. At large internuclear distances $R$, the ground electronic state corresponds to both atoms in the $2s$ state, while the excited electronic state represents a configuration with one atom in the $2p$ state as $R\rightarrow\infty$.
	
	In our next approximation, we consider only one active electron. Under this approximation, the ground electronic state corresponds to a $2s\sigma$ state and the excited state to a $2p\sigma$ state. Following Hund's case (b) representation,\cite{lefebvre2004spectra} the electronic wavefunctions in the molecular frame are expressed as 
	\begin{eqnarray}
		\label{Hund}
		\Phi_{g}\left(\boldsymbol{r},R  \right)=\varphi_{g}(r,R) Y_{00}(\hat{r}), \\
		\Phi_{e}\left(\boldsymbol{r},R  \right)=\varphi_{e}(r,R) Y_{10}(\hat{r}),
	\end{eqnarray}
	where $\varphi_{g,e}(r,R)$ represent the radial components and $Y_{00}(\hat{r})$, $Y_{10}(\hat{r})$ represent the angular components of the active electron wavefunction in the ground and excited electronic states, respectively.
	
	To couple molecules with the local EM field calculated in the laboratory frame, we expand the ro-vibrational wavefunctions in terms of normalized Wigner rotational matrices $D^{J}_{M,0}(\hat{R})$\cite{Zare1989}
	\begin{equation}
		\label{chi-expansion}
		\chi_{g,e}\left( \boldsymbol{R},t\right)=\sum_{J,M}\chi^{(g,e)}_{JM}\left(R,t\right)D^{J*}_{M,0}(\hat{R}),
	\end{equation}
	where $J$ denotes the molecular rotational quantum number and $M$ its projection onto the laboratory frame's $z$-axis. Upon substituting Eq. (\ref{chi-expansion}) into the time-dependent Schr$\ddot{\text{o}}$dinger equation and projecting onto the electronic and rotational basis functions, we obtain a set of coupled equations describing quantum ro-vibrational dynamics. The effective Hamiltonians take the form
	\begin{equation}
		\label{hamiltonian}
		\hat{H}_J^{(g,e)}=-\frac{\hbar^2}{m}\frac{\partial^2}{\partial R^2}+V_{g,e}(R)+\frac{\hbar^2}{mR^2}J(J+1),
	\end{equation}
	where $m$ represents the nuclear mass. The equations read\cite{Sukharev2017}
	\begin{eqnarray}
		\label{chi-equations}
		i\hbar\frac{\partial\chi^{(g)}_{JM}}{\partial t}=\hat{H}_J^{(g)}\chi^{(g)}_{JM}-\mu_{eg}(R)\sum_{J'M'}\Lambda_{JM}^{J'M'}(\boldsymbol{E})\chi^{(e)}_{J'M'}, \\
		i\hbar\frac{\partial\chi^{(e)}_{JM}}{\partial t}=\hat{H}_J^{(e)}\chi^{(e)}_{JM}-\mu_{ge}(R)\sum_{J'M'}\left(\Lambda_{JM}^{J'M'}(\boldsymbol{E})\right)^*\chi^{(g)}_{J'M'}.
	\end{eqnarray}
	The electronic component of the transition dipole moment is given by 
	\begin{equation}
		\label{dipole}
		\mu_{eg}(R)=\langle \varphi_{e}(r,R) | \; er \; | \varphi_{g}(r,R) \rangle.
	\end{equation}
	For all calculations, we assume the transition dipole moment is independent of $R$ and has a constant value of $\mu_{eg}(R)=15$ Debye. The rotational component of the transition dipole moment couples to the local electric field $\boldsymbol{E}$ in the laboratory frame and is denoted as $\Lambda_{JM}^{J'M'}(\boldsymbol{E})$. This component can be expressed in terms of Wigner 3-j symbols as follows\cite{Charron2006}
	\begin{equation}
		\label{3j-coupling}
		\Lambda_{JM}^{J'M'}(\boldsymbol{E})=X_{JM}^{J'M'}E_x+Y_{JM}^{J'M'}E_y+Z_{JM}^{J'M'}E_z,
	\end{equation}
	where matrices $X$, $Y$, and $Z$ read
	\begin{eqnarray}
		X_{J M}^{J' M'}=\frac{(-1)^{M}}{\sqrt{2}}\sqrt{\left(2J+1 \right)\left(2J'+1 \right)}\begin{pmatrix} J & 1 & J' \\ 0 & 0 & 0 \end{pmatrix}\times\nonumber \\
		\times\left(\begin{pmatrix} J' & 1 & J \\ M' & -1 & -M \end{pmatrix} - \begin{pmatrix} J' & 1 & J \\ M' & 1 & -M \end{pmatrix}\right), \\
		Y_{J M}^{J' M'}=i\frac{(-1)^{M+1}}{\sqrt{2}}\sqrt{\left(2J+1 \right)\left(2J'+1 \right)}\begin{pmatrix} J & 1 & J' \\ 0 & 0 & 0 \end{pmatrix}\times\nonumber \\
		\times\left(\begin{pmatrix} J' & 1 & J \\ M' & -1 & -M \end{pmatrix} + \begin{pmatrix} J' & 1 & J \\ M' & 1 & -M \end{pmatrix}\right), \\
		Z_{J M}^{J' M'}=(-1)^{M}\sqrt{\left(2J+1 \right)\left(2J'+1 \right)}\begin{pmatrix} J & 1 & J' \\ 0 & 0 & 0 \end{pmatrix} \begin{pmatrix} J' & 1 & J \\ M' & 0 & -M \end{pmatrix}.
	\end{eqnarray}
	
	In coupling Maxwell's equations to molecular quantum dynamics, the EM field is calculated in Cartesian coordinates in the laboratory frame. Within molecular regions, the quantum dynamics described by Eqs. (\ref{chi-equations}) evolve in the molecular frame using the local electric field. The significant spatial variations of the electric field necessitate position-dependent molecular frames. The quantum-mechanically calculated induced dipole moment couples back to Maxwell's equations through the polarization current. Computational efficiency is achieved through the analytical integration of the angular components $\Lambda_{JM}^{J'M'}(\boldsymbol{E})$ in the quantum dynamical equations, which facilitates transformations between laboratory and molecular frames. This analytical approach enables direct computation of molecular orientation and alignment dynamics across complex optical environments, ranging from Fabry-Perot cavities to three-dimensional plasmonic metasurfaces.
	
	\section{Ro-vibrational dynamics in lossy Fabry-Perot cavities} \label{1d-discussion}
	In this section, we discuss ro-vibrational dynamics of the ensemble of molecules placed in-between two flat metallic mirrors as depicted in Fig. \ref{fig:1}c. EM field follows classical Maxwell's equations adopted for the 1D-geometry with $x$ axis corresponding to the propagation direction of the incident field (considered here at normal incidence). The electric field is polarized along $z$ axis
	\begin{eqnarray}
		\label{1d-Maxwell}
		\frac{\partial B_y}{\partial t} &=& \frac{\partial E_z}{\partial x}, \\
		\frac{\partial E_z}{\partial t} &=& c^2\frac{\partial B_y}{\partial x}-\frac{1}{\varepsilon_0} J_z,
	\end{eqnarray}
	where the current density inside the molecular slab is written as a time derivative of the macroscopic polarization, $J_z\equiv\frac{\partial P_z}{\partial t}$, which is evaluated using constant molecular number density $n_M$ and the quantum mechanical averaged induced molecular dipole $\mu_z$ projected onto the field polarization following the mean field approximation\cite{Sukharev2011}
	\begin{equation}
		\label{P-meanfield}
		P_z=n_M \mu_z.
	\end{equation}
	Optical response of the metal is described by the Drude-Lorentz model with multiple poles
	\begin{equation}
		\label{epsilon-Au}
		\varepsilon(\omega)=1-\frac{\Omega_D^2}{\omega^2-i\Gamma_D\omega}-\sum_n\frac{\Delta\varepsilon_n\omega_p^2}{\omega^2-\omega_n^2-i\Gamma_n\omega},
	\end{equation}
	where parameters are taken from Ref.\cite{Rakic1998} accounting for the material dispersion of gold in the range of wavelengths between 200 nm and 1.9 microns. The dielectric function Eq. (\ref{epsilon-Au}) leads to the set of coupled differential equations on the current density in metal\cite{Pernice2006}
	\begin{eqnarray}
		\label{DrudeLorentz-J}
		\frac{\partial J_z^{(D)}}{\partial t}+\Gamma_D J_z^{(D)} &=& \varepsilon_0 \Omega_D^2 E_z, \\
		\frac{\partial^2 J_z^{(n)}}{\partial t^2}+\Gamma_n\frac{\partial J_z^{(n)}}{\partial t}+\omega_n^2 J_z^{(n)} &=& \varepsilon_0 \Delta\varepsilon_n \omega_n^2 \frac{\partial E_z}{\partial t}, \\
		J_z &=& J_z^{(D)}+\sum_n J_z^{(n)}.
	\end{eqnarray}
	
	We explore the one-dimensional geometry under specific assumptions. The metal slabs and molecular layer are considered to extend infinitely in the $y$ and $z$ directions while remaining finite in the $x$ direction, reducing the problem to a single spatial dimension, $x$. To maintain this symmetry, the incident EM field must excite the system at normal incidence, with the wave vector $k$ aligned along the $x$ axis, ensuring uniform interaction across any $x=$const cross-section. This restricts the wave vector components to zero in the $yz$ plane. Additionally, the initial conditions of the molecules must depend solely on the $x$ coordinate to remain consistent with the one-dimensional model, limiting any variation in molecular properties to changes along the $x$ direction.
	
	The coupled system of Eqs. (\ref{chi-equations}), (\ref{1d-Maxwell}), and (\ref{DrudeLorentz-J}) is discretized in space using Yee's spatial decomposition for the EM field\cite{Taflove2005} and evolved in time using leap-frog propagation, following the finite-difference time-domain (FDTD) methodology. The EM grid boundaries (Fig. \ref{fig:1}c) are terminated using convolution perfectly matched layers (CPMLs) to simulate open boundaries.\cite{Roden2000} Numerical convergence for EM field propagation is achieved with a spatial resolution of $\delta x_{\text{EM}}=1.5$ nm and a time step of $\delta t_{\text{EM}}=2.5$ as. The ro-vibrational wavefunctions (\ref{chi-equations}) are discretized on the internal molecular grid with a spatial resolution of $\delta R=0.08$ a.u. and propagated using the multi-surface split-operator method.\cite{Charron1998} For strong laser pulse excitation, convergence requires a maximum rotational quantum number of $J_{max}=6$. Due to the vibrational dynamics occurring on significantly slower times scales compared to the EM field, Eqs. (\ref{chi-equations}) can be updated every $0.1$ fs.
	
	Numerical convergence in modeling the self-consistent dynamics of many quantum emitters coupled through Maxwell's equations requires specific conditions. For a continuous spatial distribution of molecules, the molecular number density $n_M$ and the spatial discretization of Maxwell's equations $\delta x_{\text{EM}}$ must satisfy the following inequality
	\begin{equation}
		\label{density-convergence}
		n_M \ll \frac{1}{\delta x_{\text{EM}}^3}.
	\end{equation}
	This condition ensures consistency of physical properties across neighboring spatial positions. While individual molecules can be simulated as discrete delta functions ($n_M=\frac{1}{\delta x_{\text{EM}}^3}$), their spatial separation must be appropriately discretized and free of other materials.\cite{Neuhauser2007,Lopata2009} For scenarios with relatively low electron excitation, numerical convergence typically requires approximately 10 spatial points between molecules.
	
	Our model is built upon the semi-classical Ehrenfest approach, which describes the interaction between classical electromagnetic fields and quantum systems. While this framework has proven valuable for many applications, it notably lacks the ability to capture spontaneous emission - a fundamental quantum phenomenon crucial for a complete description of light-matter interactions. Although incorporating spontaneous emission into the Ehrenfest framework presents significant theoretical and computational challenges, recent promising developments\cite{Roel1,Roel2} have demonstrated potential methods for extending Ehrenfest dynamics to include these quantum effects. These advances represent an important step toward developing more comprehensive and reliable quantitative models for polaritonic chemistry, where accurate representation of quantum phenomena is essential for predicting and understanding chemical processes modified by strong light-matter coupling
	
	We begin by analyzing the linear response of molecules within a one-dimensional cavity. Figure \ref{fig:1}d shows the linear absorption spectrum for three configurations: an empty cavity (black line), a cavity containing molecules treated as two-level systems without internal degrees of freedom (red line), and a cavity with molecules described by PESs (\ref{PESs}) incorporating full molecular complexity (blue line). First, we calculate the EM energy flux on the input (reflection, $R$) and output (transmission, $T$) sides of the system. Absorption is evaluated as $1-T-R$.
	
	The metal slab separation is tuned to align the EM fundamental mode with the $0-0$ transition at 1.85 eV (vertical dashed line). The simulations start with molecules in their ground vibrational state of the electronic ground state $|g\rangle$, with rotational quantum number $J=0$. The lower polariton (LP) appears clearly in both molecular cases and shows minimal sensitivity to ro-vibrational structure. For molecules without vibrations, the observed Rabi splitting is 111 meV, representing the energy separation between polaritonic states under strong coupling. When ro-vibrational degrees of freedom are included, the upper polariton region develops a complex structure with several peaks displaying Fano lineshapes, indicating significant interference effects between electronic and ro-vibrational states.
	
	\begin{figure}
		\centering
		\includegraphics[width=1.0\linewidth]{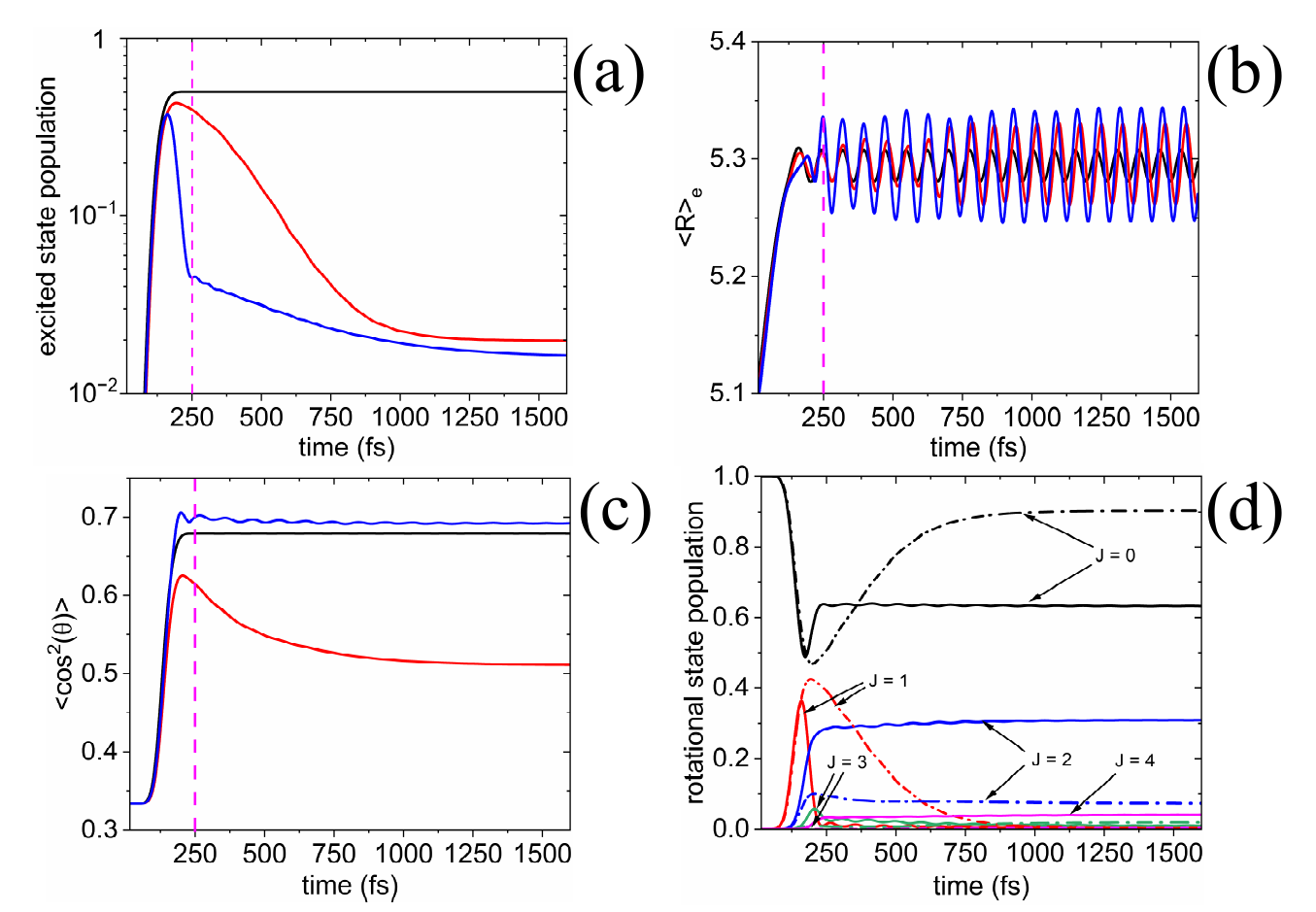}
		\caption{Results for molecules being pumped by a strong 250 fs laser pulse resonant with the $0-0$ transition $\Omega_{\text{mol}}=1.85$ eV. Magenta vertical dashed line in panels (a) - (c) indicates the end of the incident pulse. All plots show data after averaging over the molecular ensemble. (a) Excited electronic state population (on a log scale) as a function of time in fs. Black line shows results for a single molecule in vacuum, red line is for the slab of molecules outside the cavity, blue line is for the molecules inside the resonant cavity. (b) Active coordinate of the excited molecules as a function of time in fs. (c) The degree of molecular alignment $\left\langle \cos^2(\theta) \right\rangle$.  The color scheme for both (b) and (c) is the same as in panel (a). (d) Populations of first 5 rotational states as functions of time, dash-dotted lines show populations for molecules outside the cavity and solid lines are for the molecules inside the cavity. The number density of molecules is $n_M=10^{26}$ m$^{-3}$.}
		\label{fig:2}
	\end{figure}
	
	Fig. \ref{fig:2} examines the relaxation dynamics of molecules following the resonant excitation. We use a 250-fs long incident pulse with a peak amplitude of 0.1 V/nm centered at $0-0$ transition frequency of 1.85 eV. The time envelope of the pulse is taken in the form of the Blackman-Harris function.\cite{Oppenheim1999} Fig. \ref{fig:2}a presents the evolution of the excited electronic state population for three distinct cases: single molecule in vacuum driven by the incident pulse only without any interaction with environment (black line), the slab of molecules in free space (red line), and the molecules inside the resonant cavity (blue line). The excitation for a single molecule in vacuum reaches 0.5 and stays flat once the pulse is off as expected since the molecule is not interacting with itself nor the relaxation occurs due to spontaneous emission, which is not accounted for. The ensemble of molecules behaves noticeably different. When no cavity is present, the excitation is slightly lower than that of a single molecule, but the relaxation is clearly visible. This is attributed to the energy transfer between molecules in free space, which is partially radiated to the far-field and thus lost. When a resonant cavity is present, an additional channel for energy loss is introduced through Ohmic losses in the metal. The relaxation however is vastly different from that in free space. The dynamics of molecules in the cavity peaks at 0.38, compared to 0.43 in free space. The discrepancy arises because the incident field first encounters the metal slab, losing some energy due to Ohmic losses in metal. Maximum excitation in the cavity occurs earlier, at 160 fs, compared to 195 fs in free space. Notably, the fast transient relaxation observed in free space is absent in the cavity scenario. Moreover, several oscillations appear after the incident pulse in the cavity setup (see Fig. \ref{fig:2}a, blue line). These oscillations are indicative of the Rabi splitting observed in the linear absorption (Fig. \ref{fig:1}d) and correspond to the energy exchange between the cavity mode and molecular excitation. This behavior underscores the complex interactions and energy dynamics facilitated by the cavity environment under the strong coupling.
	
	Figs. \ref{fig:2}b and \ref{fig:2}c respectively present the vibrational and rotational dynamics following strong resonant pulse excitation. The average internuclear separation for excited molecules, $\langle R \rangle_{\text{e}}$ oscillates with a period of 75 fs after the pulse, corresponding to the vibrational period in the excited electronic state, $T_e=\frac{\pi}{a_e}\sqrt{\frac{m}{D_e}}$. The vibrational dynamics follow a rapid vertical electronic transition, as evidenced by the monotonic increase in $\langle R \rangle_{\text{e}}$ at early times, which starts at $R_g$ and reaches $R_e$ around 160 fs (see Eq. (\ref{PESs})). Notably, the amplitude of vibrations depends on the environment, being highest for molecules in the cavity. In one-dimensional geometries, molecular alignment occurs along the local electric field polarization. A useful indicator for examining rotational dynamics is the degree of alignment, defined as $\langle \cos^2(\theta) \rangle$, where $\theta$ is the angle between the molecular axis and the field polarization.\cite{Stapelfeldt2003} For the initial state with $J=0$, $\langle \cos^2(\theta) \rangle = \frac{1}{3}$. From Fig. \ref{fig:2}c shows that $\langle \cos^2(\theta) \rangle$ is highest for molecules in the cavity, oscillating with the vibrational period once the pump is off. Observed oscillations indicate the fact that vibrations influence the rotational dynamics. Although the oscillation amplitude gradually decreases, the degree of alignment remains around 0.7 for an extended period, indicating that molecules in a resonant cavity tend to remain aligned with the cavity field long after the pump ceases. For a single molecule in a vacuum, $\langle \cos^2(\theta) \rangle$ is slightly lower than for molecules in the cavity and remains constant due to the lack of environmental interaction. The alignment of a slab of molecules outside the cavity dissipates once the pump is removed.
	
	With our approach, it is possible to directly access the dynamics of each rotational state. Figure \ref{fig:2}d illustrates the populations of several rotational states during and after excitation. We examine the rotational dynamics of molecules inside the cavity (solid lines) and outside the cavity (dash-dotted lines). Initially, at $t=0$, all molecules are in the $J=0$ state in the ground electronic state. The electronic transitions between the two states, $| g \rangle$ and $| e \rangle$, result in a change in the rotational quantum number by 1 due to the symmetry of the electronic states, which is reflected in the matrix elements (see Eq. (\ref{3j-coupling})). For both cases molecules inside and outside the cavity exhibit rotational excitation due to the pump. However, after the pulse, molecules in free space decay to the ground electronic state and return to $J=0$. In contrast, molecules within the cavity relax to $J=2$, which explains why $ \langle \cos^2(\theta) \rangle$ remains high over longer periods. Additionally, there is a noticeable population in the $J=4$ state, indicating the occurrence of multiple electronic excitations ($\Sigma_g^+(J=0) \rightarrow \Sigma_u^+(J=1) \rightarrow \Sigma_g^+(J=2) \rightarrow \Sigma_u^+(J=3) \rightarrow \Sigma_g^+(J=4)$) that result in rotational pumping. This behavior underscores the influence of the cavity on rotational dynamics, facilitating higher rotational states and sustained alignment. 
	
	Through extensive numerical simulations exploring various cavity geometries, coupling strengths, and molecular parameters, we have consistently observed this rotational pumping phenomenon whenever strong coupling conditions are satisfied. The EM environment, when resonant with a given electronic transition following an intense resonant pump, consistently leads to electronic relaxation to a lower electronic state but with a higher rotational quantum number. This robust behavior appears to be a fundamental feature of strongly coupled molecule-cavity systems rather than an artifact of specific parameters or geometries. The underlying mechanism for this preferential relaxation pathway remains unclear, as does the precise reason for the sustained molecular alignment in resonant cavities. These observations raise intriguing questions about the nature of energy redistribution in strongly coupled quantum systems and the interplay between electronic and rotational degrees of freedom in resonant EM environments.

	\begin{figure}
		\centering
		\includegraphics[width=1.0\linewidth]{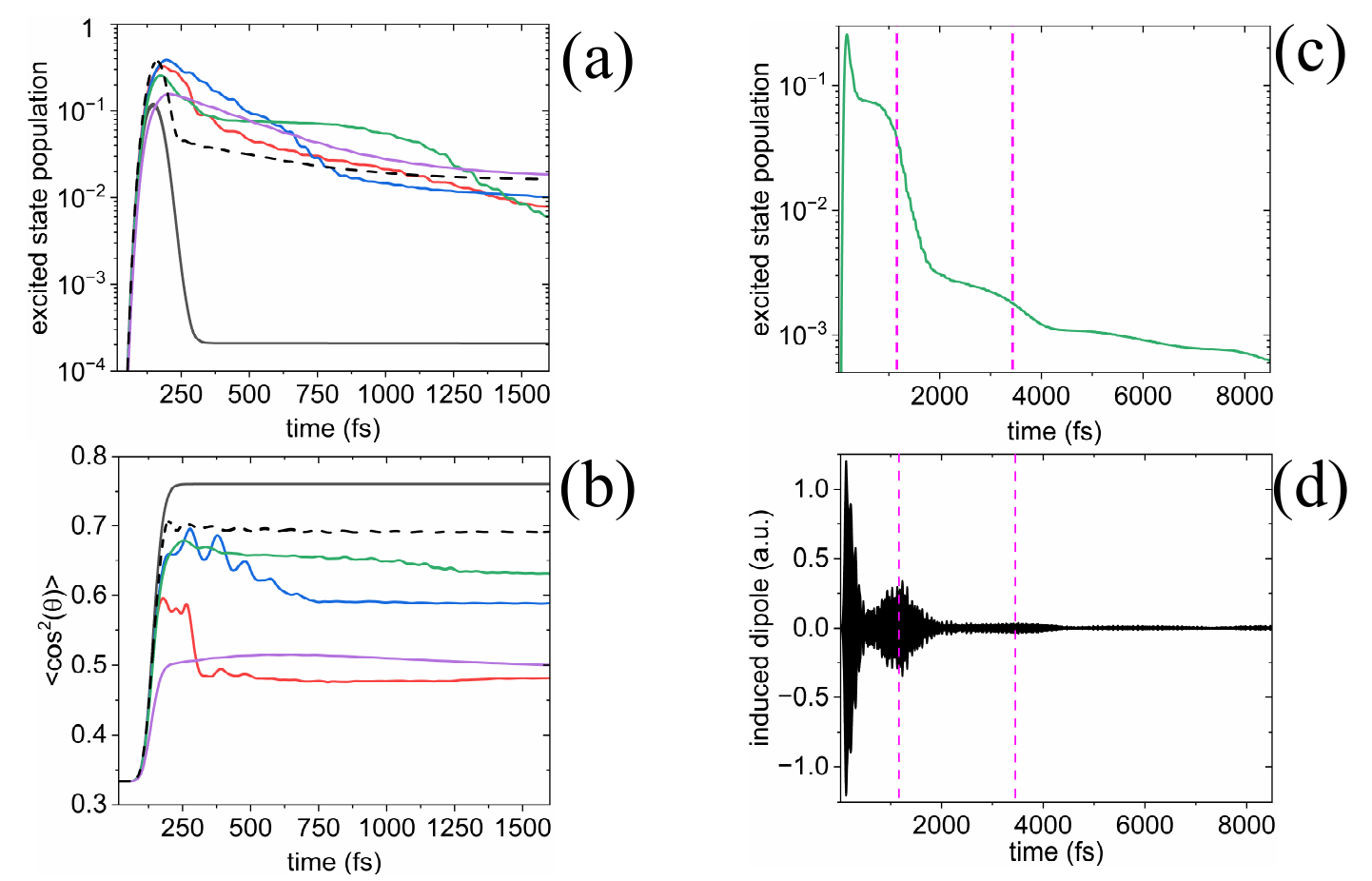}
		\caption{Results for the molecules pumped at frequencies (see Fig. \ref{fig:1}d). In all panels black solid lines correspond to the pump centered at $\Omega_0$ (lower polariton), black dashed lines – at $\Omega_{\text{mol}}$, red lines – at $\Omega_1$ (first absorption maximum to the right of the $0-0$ transition $\Omega_{\text{mol}}$), blue lines – at $\Omega_2$, green lines – at $\Omega_3$, and magenta lines – at $\Omega_4$. Panel (a) shows dynamics of the excited electronic state population (on a log scale). Panel (b) shows the degree of alignment, $\langle \cos^2(\theta) \rangle$. Panel (c) shows the extended dynamics for the excited electronic state population pumped at $\Omega_3$. Two vertical magenta dashed lines indicate two maxima of the induced molecular dipole after the pump. Panel (d) presents the induced molecular dipole for the pump at $\Omega_3$. The number density of molecules is $n_M=10^{26}$ m$^{-3}$.}
		\label{fig:3}
	\end{figure}
	
	Fig. \ref{fig:3} explores the ro-vibrational dynamics under different excitation frequencies indicated in Fig. \ref{fig:1}d. The lower polariton excitation exhibits the fastest relaxation (panel (a)) and achieves the highest degree of alignment (panel (b)). At other frequencies, the population and alignment dynamics display highly complex time dependencies, with oscillations corresponding to both the vibrational period and Rabi cycling, due to strong coupling with the cavity mode. Intriguing dynamics is observed at $\Omega_3$ (green lines), where electronic relaxation is nearly halted between 400 fs and 800 fs. Similar brief stabilization periods are also evident for pumps at $\Omega_1$ (red lines) and $\Omega_2$ (blue lines). Our extensive numerical analysis across various pump frequencies and amplitudes reveals that molecules in resonant cavities exhibit relaxation stabilization when excited at specific frequencies, while molecules in free space show monotonic decay in both electronic excitation and alignment following the pump. This stabilization is most pronounced when the pump frequency exceeds the 0-0 transition, as it excites a wavepacket comprising several ro-vibrational states in the excited electronic state.
	
	Figs. \ref{fig:3}c and \ref{fig:3}d illustrate multiple stabilization plateaus in the electronic excitation dynamics and their correlation with the induced molecular dipole behavior. The dipole exhibits several bursts following the initial pump excitation, with two prominent peaks (marked by vertical magenta dashed lines) corresponding to electronic excitations relaxing after stabilization plateaus. This phenomenon occurs when the collective macroscopic molecular polarization experiences revivals, alternating between in-phase and out-of-phase states with the resonant cavity mode. When the molecular polarization is out of phase, it becomes detuned from the cavity mode resonance, reducing energy exchange and creating stabilization plateaus. Conversely, when in phase, rapid relaxation occurs through enhanced coupling to the electromagnetic mode. The duration of each plateau is defined by the pump frequency and the ro-vibrational states it excites.
	
	\begin{figure}
		\centering
		\includegraphics[width=1.0\linewidth]{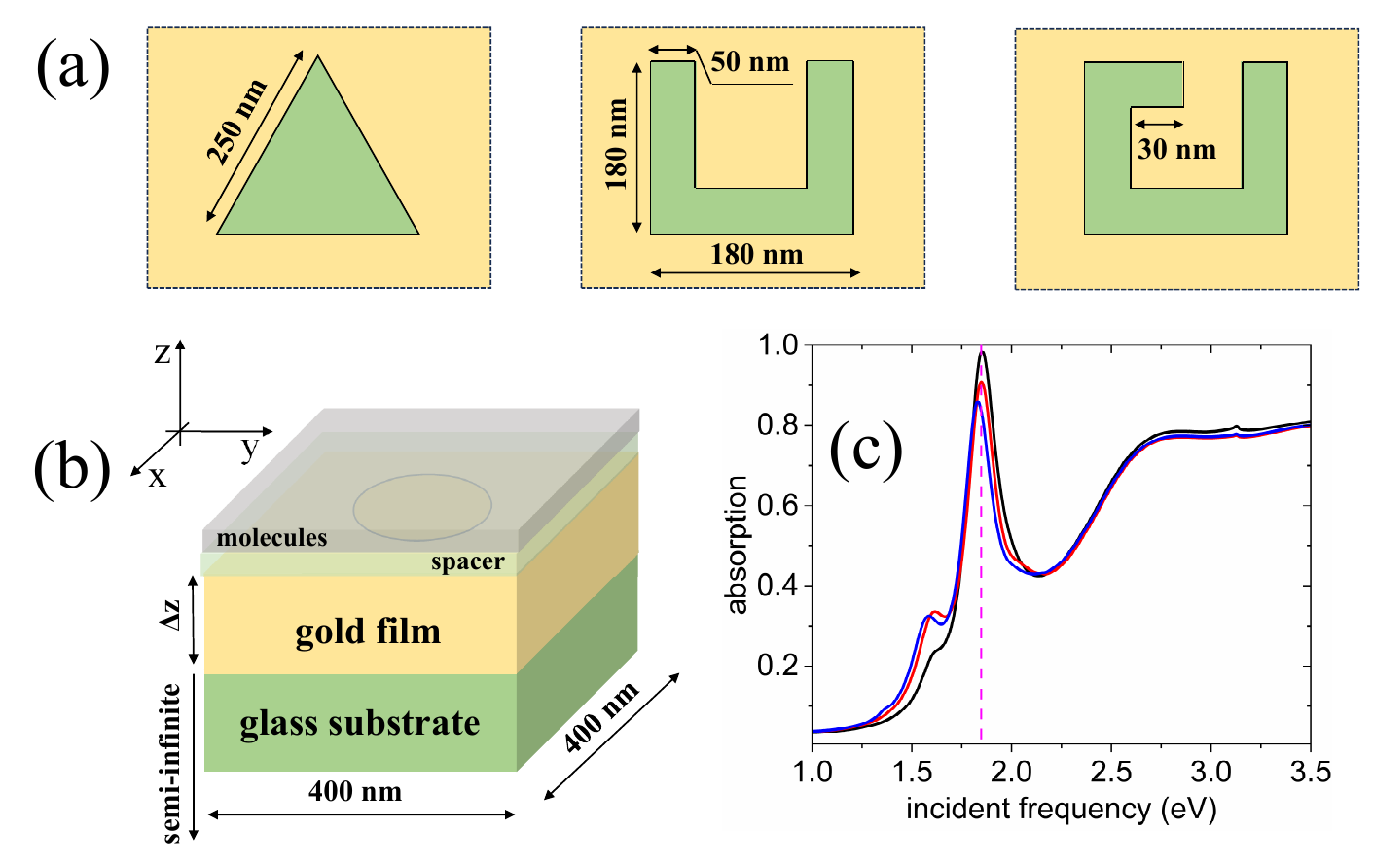}
		\caption{Three-dimensional setup showing geometries of the plasmonic metasurfaces considered. All surfaces are periodic arrays of nano-holes of three distinct geometries: (a) (from left to right) triangular holes (TH), holes in the shape of a split-ring resonator (SRR), and chiral SRRs with one of the arms twisted (CSRR) (b) The unit cell for three-dimensional EM simulations. Molecular layer is 11 nm thin. It is placed on top of a 10 nm thick dielectric spacer. The metasurface is on top of a semi-infinite glass substrate. (c) Linear absorption spectra for metasurfaces (simulations without molecules): black line shows data for the array of triangular holes, red line is for the split-ring resonators array, and blue line is for the chiral split-ring resonator array. The arrays parameters are optimized to have the local plasmon mode centered at 1.85 eV.}
		\label{fig:4}
	\end{figure}
	
	\section{Three-dimensional plasmonic metasurfaces} \label{metasurfaces}
	In this section, we consider three-dimensional plasmonic interfaces composed of periodic arrays of holes and investigate the ro-vibrational dynamics of diatomic molecules under strong coupling conditions with respect to a chosen surface plasmon-polariton (SPP) mode. The geometries under consideration are depicted in Figure \ref{fig:4}. These include triangular holes (THs), holes shaped as inverted $\pi$, commonly referred to as split-ring resonators (SRRs), and SRRs with one arm twisted to induce chiral EM fields, known as chiral SRRs (CSRRs). The spatio-temporal dynamics of EM radiation follows Maxwell's equations in three-dimensions
	\begin{eqnarray}
		\label{3d-Maxwell}
		\frac{\partial \textbf{B}}{\partial t} &=& -\nabla \times \textbf{E}, \\
		\frac{\partial \textbf{E}}{\partial t} &=& c^2 \nabla \times \textbf{B}-\frac{1}{\varepsilon_0} \textbf{J}.
	\end{eqnarray}
	The current density $\textbf{J}$ in metal is evaluated using a three-dimensional analogue of Eq. (\ref{DrudeLorentz-J}). The Maxwell equations, along with equations on the current density, are discretized in space and propagated in time using the FDTD method. Due to the high computational demands of three-dimensional dynamics, we employ a three-dimensional domain decomposition (3D-DD) method and split the simulation domain into $12\times 12\times 16 = 2304$ cubes, each representing a single processor.\cite{Sukharev2023} The discretized equations are propagated in time in parallel using message passing interface (MPI) subroutines, ensuring proper calculations of the terms $\nabla \times \textbf{E}$ and $\nabla \times \textbf{B}$ at all six faces of each computational cube. This parallel approach has been proven to be very efficient.\cite{Sukharev2023} In all simulations, we maintain the same spatial resolution of $\delta x_{\text{EM}}=1.5$ nm and time step of $\delta t_{\text{EM}}=2.5$ as in the one-dimensional problem. For the chosen geometry each of 2304 processors is carrying 37752 spatial points resulting in total of $8.7\times10^{7}$ points. The system is excited by a plane wave propagating in the negative $z$ direction (from top to bottom, see Fig. \ref{fig:4}b) with the incident field polarization in the $xy$ plane. The upper and lower boundaries in the $z$ direction are terminated using CPMLs, and the boundaries in any $xy$ plane have periodic boundary conditions. The detection planes are paralleled to the metasurface and placed 600 nm away from the structure on both input (detecting reflection, $R$) and output (detecting transmission, $T$) sides. Linear absorption calculations are the same as for Fabry-Perot geometry: the absorption is calculated using the equation $1-T-R$.
	
	First, we consider metasurfaces without molecules. The geometrical parameters of the nanoholes for each configuration depicted in Figure \ref{fig:4}a are adjusted so that the localized surface plasmon-polariton (SPP) mode has a similar linewidth for each geometry and is centered at $\Omega_{\text{mol}} = 1.85$ eV. Figure \ref{fig:4}c shows the linear absorption spectra for the three arrays without molecules. We note that a small peak around 1.5 eV is identified as a Bragg plasmon resonance, which is associated with the periodicity of the metasurfaces. Due to its spatially delocalized nature, this mode has a significantly lower mode volume compared to the SPP mode at 1.85 eV, making it challenging to achieve strong coupling with a thin molecular layer. One possible workaround is to consider periodic arrays of nanoparticles instead of holes, as these can sustain a Bragg plasmon mode in the space surrounding the nanoparticles. However, this mode still possesses a larger volume compared to an SPP. Therefore, we focus on the optics of the SPP mode and its EM properties. By concentrating on the SPP mode, we aim to understand its potential for strong coupling and its ability to enhance collective molecular interactions through localized EM fields.
	
	\begin{figure}
		\centering
		\includegraphics[width=1.0\linewidth]{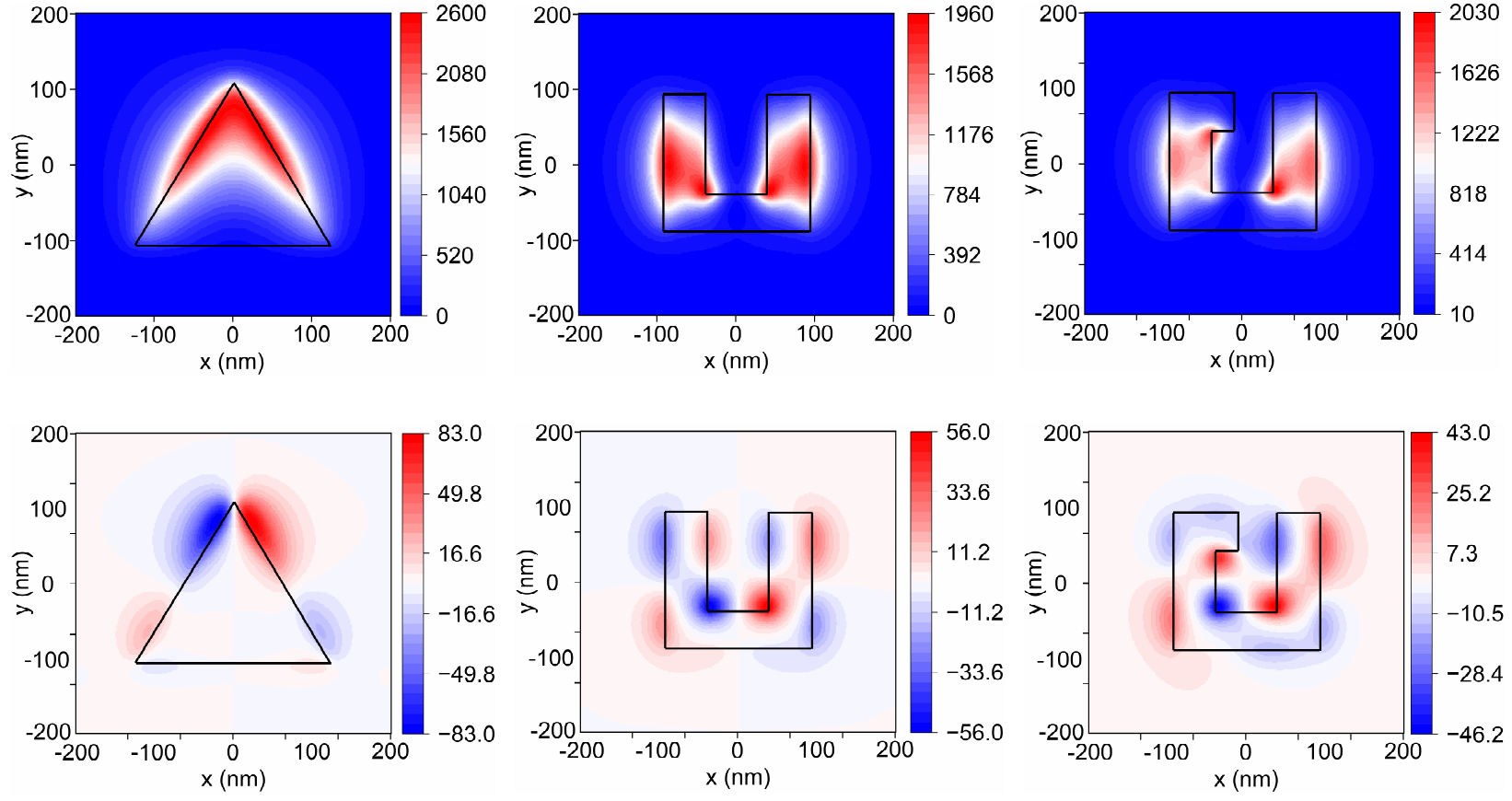}
		\caption{(simulations without molecules) Steady-state spatial distributions of the EM intensity (upper row) and the optical chirality density (lower row) evaluated at the plasmon resonance of 1.85 eV. The detection plane is 16 nm above the metal surface. The units of the intensity are given in terms of the enhancement relative to the incident field intensity. The chirality density is normalized to that of the right circular polarized field in vacuum.}
		\label{fig:5}
	\end{figure}
	
	Although the linear absorption appears to be the same for all three configurations, the local EM intensity distributions are vastly different. This is explored in Figure \ref{fig:5}, where we examine both the intensity and its polarization properties. The detection plane is positioned 16 nm above the metasurface in order to scrutinize local EM field that will be driving molecules as discussed below. The TH array exhibits the highest intensity enhancement close to the upper corner of the hole. The SRR array shows the hot spots localized on both arms down the bottom. Enhancement induced by the CSRRs is primarily near the top left and the bottom right corners. In addition to the observed local field enhancement obviously playing a crucial role in the strong coupling with molecules, the near-field polarization must have a significant impact on the molecular orientation and alignment dynamics. To explore near-field polarization properties of the SPP mode, we calculate the optical chirality density $C$ for each configuration using the conventional expression\cite{Tang2010}
	\begin{equation}
		\label{chirality}
		C=\frac{1}{2}\varepsilon_0\operatorname{Im}\left (\textbf{E}\cdot\textbf{B}^* \right).
	\end{equation}
	
	The bottom row in Fig. \ref{fig:5} shows the chirality density distributions for each metasurface configuration. Despite the geometric symmetry of both TH and SRR arrays, their near-field exhibits distinct spatial regions of EM fields with opposite handedness (left and right circular polarization). The TH configuration demonstrates two prominent regions of opposite chirality coinciding with areas of maximum field enhancement. These regions are characterized by high chirality contrast and well-defined boundaries, suggesting possible strong chiral light-matter interactions at these locations.
	
	The SRR array, while also exhibiting chiral behavior, shows a markedly different spatial distribution. It features smaller localized spots of alternating chirality, typically spanning approximately 50 nm, with notably reduced chirality contrast compared to the TH array. This reduced contrast suggests weaker chiral light-matter interactions, though the finer spatial distribution might be advantageous for applications requiring higher spatial resolution.
	
	We note that both TH and SRR configurations show no preference between left and right circular excitations in the far-field region due to the metasurfaces' symmetry. The local chirality hotspots observed in Fig. \ref{fig:5} are confined to the near-field region. This can be understood by considering a point near a triangular nano-hole: when placed asymmetrically relative to the triangle's symmetry axis, the hole's edges appear asymmetric from that point, thus inducing a local chiral response. This chiral effect vanishes when observations are made at distances significantly greater than the characteristic size of the holes. 
	
	The CSRR geometry presents the most intriguing chirality pattern among the three configurations. It displays extended regions of chirality that form a distinctive counterclockwise swirling pattern, contrasting with the clockwise orientation of the physical CSRR structures. This counterintuitive relationship between the structural and optical chirality highlights the complex nature of chiral light-matter interactions in plasmonic systems and suggests potential applications in selective enhancement of circular dichroism or chiral molecular sensing.
		
	\section{Ro-vibrational dynamics at plasmonic metasurfaces under strong coupling conditions} \label{3D-simulations}
	We now add a molecular layer comprised of diatomic molecules discussed earlier. The geometry is schematically depicted in Fig. \ref{fig:4}b. Molecules are placed on a 10 nm thin dielectric spacer separating them from the metasurface. For the spatial resolution used to discretize Maxwell's equations (\ref{3d-Maxwell}), the total number of points carrying molecules is 487872, with each grid point evolving its own set of coupled Schr$\ddot{\text{o}}$dinger equations (\ref{chi-equations}). This makes the total number of floating-point operations for each processor highly inhomogeneous in the 3D-DD parallelization scheme. 
	
	\begin{figure}
		\centering
		\includegraphics[width=1.0\linewidth]{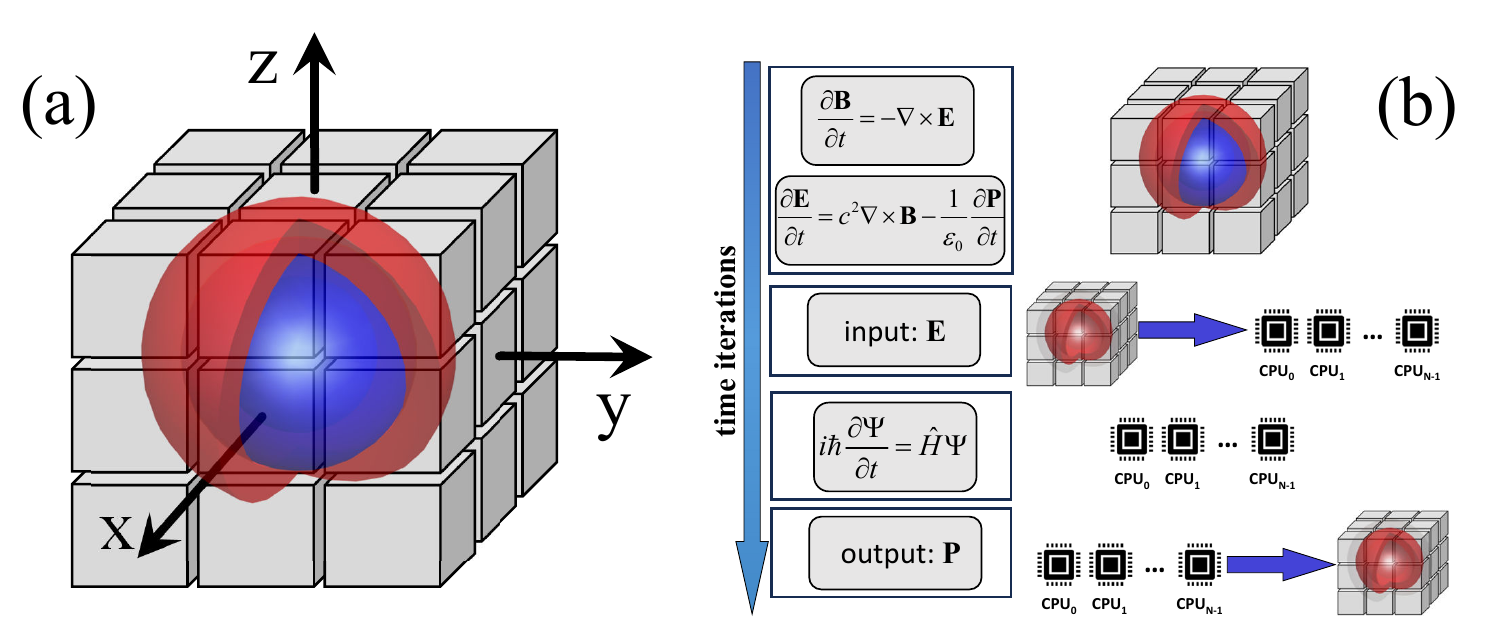}
		\caption{Schematic diagram illustrating the parallelization method, namely the molecular mapping allowing efficient three-dimensional propagation of coupled Schr$\ddot{\text{o}}$dinger-Maxwell equations for a large set of molecules.\cite{Sukharev2023} (a) The 3D-DD parallelization strategy for a hypothetical core-shell nanoparticle, where blue regions correspond to metal and red indicate regions with molecules. (b) The block diagram of the time propagation of Schr$\ddot{\text{o}}$dinger-Maxwell equations employing the molecular mapping when the quantum dynamics is involved.}
		\label{fig:6}
	\end{figure}
	
	To illustrate the major numerical bottleneck when implementing parallel simulations of the coupled Schr$\ddot{\text{o}}$dinger-Maxwell equations, we explore a hypothetical geometry shown in Fig. \ref{fig:6}a. Consider a core-shell nanoparticle with a metal core and a thin molecular shell. Each cube shown in Fig. \ref{fig:6}a represents a single processor. Those that are gray propagate Maxwell's equations in vacuum. Processors marked with the blue color integrate Maxwell's equations and corresponding current density equations (\ref{DrudeLorentz-J}). The latter are not too computationally demanding and thus the computational burden in the gray and blue spatial areas is approximately the same. However, when one adds molecules and considers their internal quantum dynamics beyond simple Bloch equations, the processors marked with red color are computationally overloaded compared to the gray and the blue ones. The more complex the molecular model employed, the more unbalanced the computational load of processors becomes.
	
	To achieve processor load balancing, we implement the molecular mapping method.\cite{Sukharev2023} Fig. \ref{fig:6}b illustrates the block scheme of this methodology. The propagation of Maxwell's equations proceeds through a conventional 3D-DD map. The method begins with a precalculation of the total grid points in the molecular layer, followed by an equal distribution of molecules across all available processors. This distribution ensures each processor advances quantum dynamics with equivalent computational load. To integrate this approach with the spatial decomposition of the EM field, we assign the local electric field that drives a specific molecule to the processor handling its dynamics. Upon calculation of the induced molecular dipole, the data is mapped back to the 3D-DD map. The efficiency of this approach depends on the relative time scales of two processes: (1) the transfer of electric field data from the original grid to all processors and the subsequent return of updated induced dipoles to the 3D-DD map, and (2) the computation of quantum dynamics. For the molecular model in this work, the molecular mapping method reduces execution times of our codes by at least an order of magnitude. The method's efficiency is even more evident for simulations with many rotational states involved. For instance, when considering strong pulse excitations with parameters used for Fabry-Perot cavities, the convergence is achieved with 7 rotational states ($J_{\text{max}}=6$). This necessitates propagating Eqs. (\ref{chi-equations}) on $J_{\text{max}}\times \left ( J_{\text{max}}+1\right)=42$ coupled PESs due to the centrifugal part in Eq. (\ref{hamiltonian}). For the spatial resolution of 1.5 nm and the typical molecular number density of $10^{26}$ m$^{-3}$ the total number of molecules inside the molecular slab in Fig. \ref{fig:4}b is 176,000.
	
	\begin{figure}
		\centering
		\includegraphics[width=1.0\linewidth]{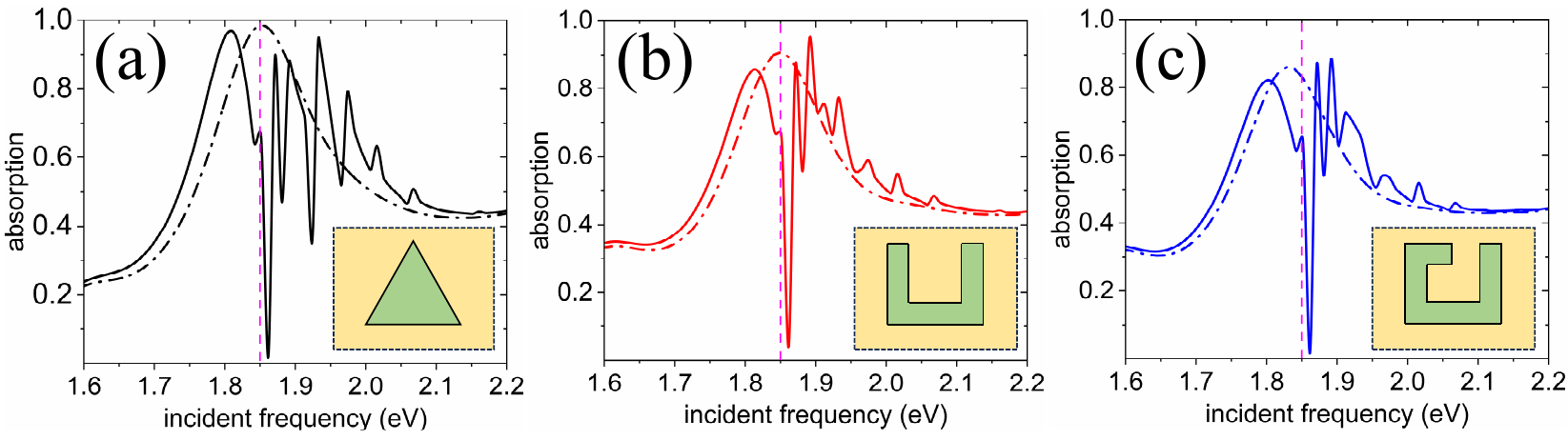}
		\caption{Linear absorption spectra for three metasurfaces with molecules: (a) TH holes, (b) SRR holes, (c) CSRR holes. Dash-dotted lines in each panel show absorption without the molecular layer and solid lines present absorption with molecules. Molecular parameters are the same as in one-dimensional simulations. Vertical magenta dashed lines indicate the $0-0$ transition frequency $\Omega_{\text{mol}}$. The number density of molecules is $n_M=10^{26}$ m$^{-3}$.}
		\label{fig:7}
	\end{figure}
	
	Figure \ref{fig:7} displays linear absorption spectra for three distinct metasurface configurations coupled to the molecular layer (geometric arrangement shown in Fig. \ref{fig:4}b). Using the same molecular parameters as employed in the Fabry-Perot cavity analysis, the spectra exhibit characteristic polariton state formation. Each configuration manifests a lower polariton branch centered at $\Omega_{\text{LP}}=1.8$ eV, accompanied by a series of sharp spectral features at frequencies exceeding the molecular transition frequency $\Omega_{\text{mol}}$. The high-frequency regime reveals ro-vibrational structure with significantly broadened spectral peaks, notably lacking the characteristic Fano lineshapes observed in one-dimensional cavities. This spectral broadening originates from the highly inhomogeneous spatial distribution of the EM field associated with the SPP mode, as illustrated in Fig. \ref{fig:5}. The spatially varying field amplitude $|\mathbf{E}(\mathbf{r},\omega)|$ drives molecular transitions with position-dependent coupling strengths $g(\mathbf{r})=\mathbf{\mu}\cdot\mathbf{E}(\mathbf{r},\omega)/\hbar$, where $\mathbf{\mu}$ denotes the molecular transition dipole moment. Additionally, the molecular orientation dynamics exhibits significant spatial variation due to the local EM field polarization containing multiple domains of distinct helicity. This dual inhomogeneity in the local field amplitude and its polarization state $\boldsymbol{\epsilon}(\mathbf{r})$, characterized by the spatial distribution of optical chirality $C$, effectively suppresses EM interference effects in the absorption spectrum.
	
	\begin{figure}
		\centering
		\includegraphics[width=0.5\linewidth]{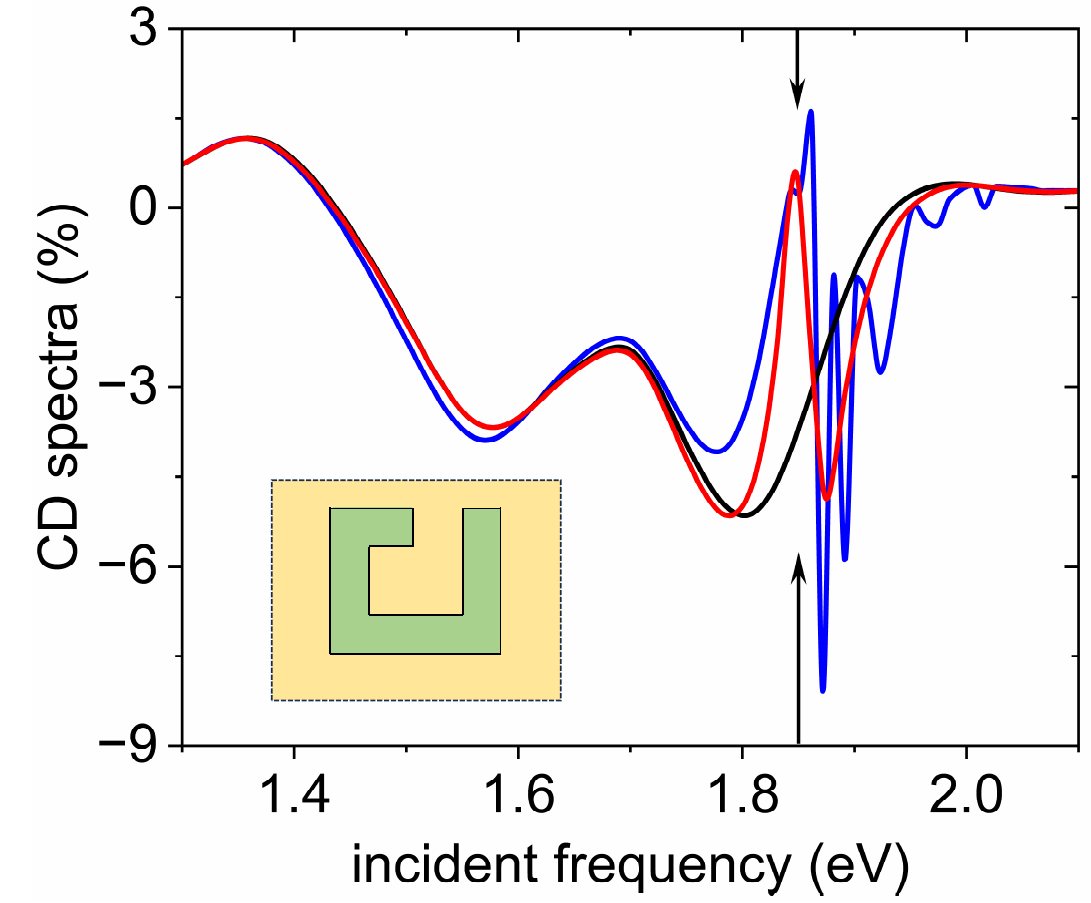}
		\caption{CD spectra simulated for the chiral metasurface, CSRR. Black line shows results for the bare metasurface without molecuels, red solid line shows CD spectrum for two-level emitters (the number density used here is $8\times10^{25}$ m$^{-3}$) and the blue line presents results for the CSRR metasurface with molecules with ro-vibrational degrees of freedom (the molecular number density is $10^{26}$ m$^{-3}$). The vertical arrows indicate the $0-0$ transition frequency $\Omega_{\text{mol}}$. The density of two-level emitters is used here as an adjustable parameter to match the Rabi splitting obtained in Fig. \ref{fig:7}c. }
		\label{fig:8}
	\end{figure}
	
	We examine the influence of molecular ro-vibrational degrees of freedom on the circular dichroism (CD) under strong coupling conditions. The metasurface configuration with chiral SRR holes exhibits asymmetric regions of enhanced field intensity ("hot spots") with varying handedness, as illustrated in Fig. \ref{fig:5}. Figure \ref{fig:8} presents CD spectra for three configurations of the CSRR metasurface: (1) the metasurface without molecules, (2) with molecules with frozen ro-vibrational degrees of freedom (represented as two-level emitters, as discussed in relation to Fig. \ref{fig:1}d), and (3) with molecules with active ro-vibrational modes. The CD spectrum is defined as
	\begin{equation}
		\label{CD}
		\text{CD}=\frac{\alpha_L-\alpha_R}{\alpha_L+\alpha_R},
	\end{equation}
	where $\alpha_{L,R}$ denotes the linear absorption evaluated for left (right) circularly polarized incident probe fields. The bare CSRR holes exhibit a predominantly negative CD signal near the SPP resonance, characterized by two distinct minima centered at 1.6 eV and 1.8 eV, and a single broad maximum at 1.35 eV. These spectral features correspond to the Bloch plasmon modes discussed in relation to Fig. \ref{fig:4}c. The introduction of the molecular layer substantially modifies the CD response near the molecular transition frequency $\Omega_{\text{mol}}$. The two-level molecular model produces a derivative-like lineshape in the CD spectrum. Molecules with rotational and vibrational degrees of freedom generate additional sets of narrow spectral features, analogous to those observed in the absorption spectrum shown in Fig. \ref{fig:7}. It should be noted that the initial conditions for molecules have no pre-defined handedness, and thus the observed CD response in the far-field is due to the strong modification of the local EM field chirality by the molecules under strong coupling conditions. The observed substantial changes in CD spectra induced by molecular adsorption present an intriguing phenomenon that remains not fully understood, and is subject to ongoing scientific discussion due to its potential applications in enantiomer detection and chiral sensing. Particularly remarkable is the complete reversal of CD contrast when molecules are present (blue line in Fig. \ref{fig:8}): the system transitions from right-handed dominance in the absence of molecules to left-handed dominance upon molecular adsorption. This dramatic change in chiroptical response suggests a complex interplay between molecular and plasmonic chirality that could be leveraged for highly sensitive molecular detection and characterization techniques. We anticipate that local molecular CD properties can be altered by placing molecules at given near-field hot spots with desired chirality.\cite{Pfaffenberger2023} 
	
	\begin{figure}
		\centering
		\includegraphics[width=0.9\linewidth]{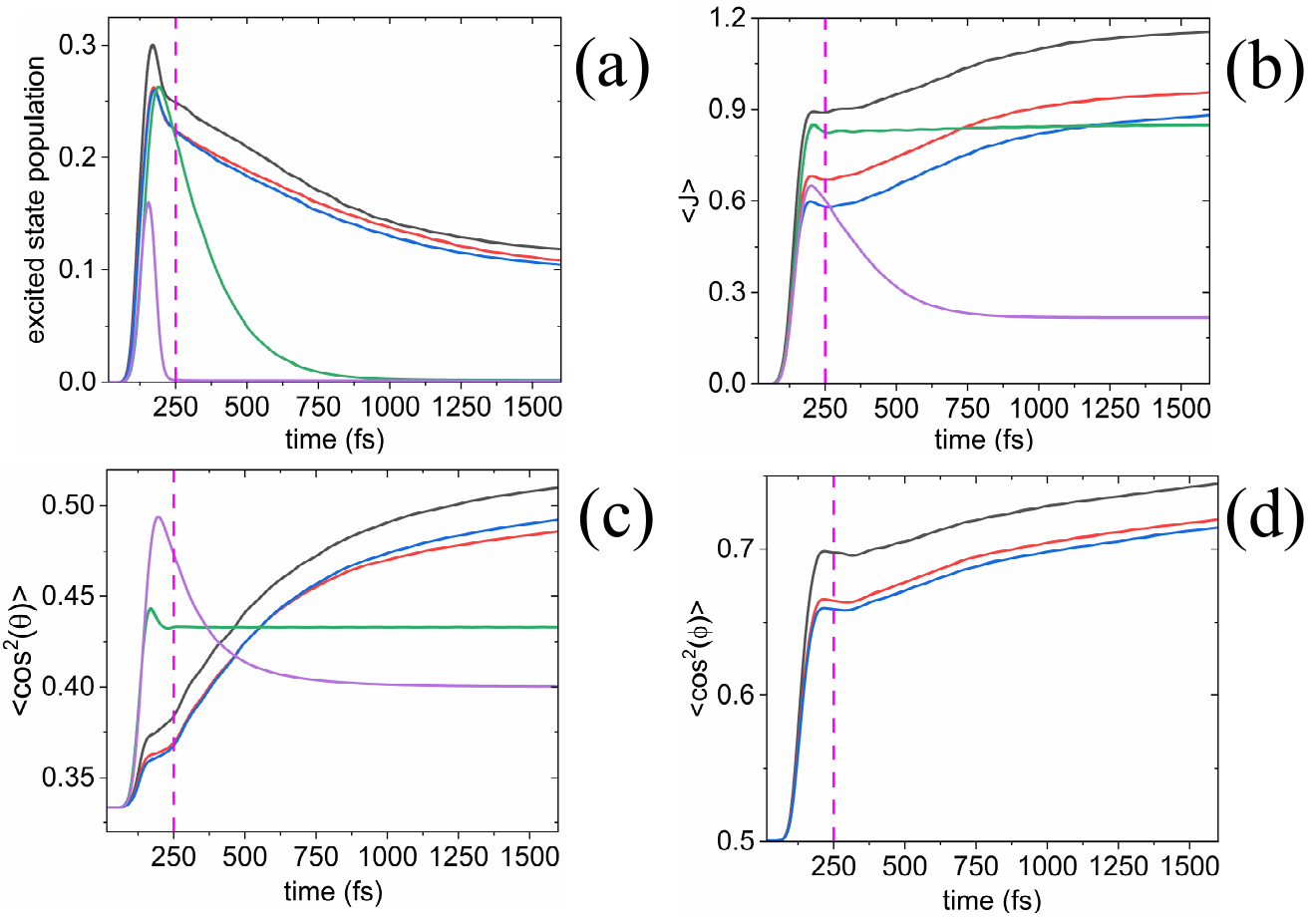}
		\caption{Three-dimensional simulations of molecular dynamics coupled to plasmonic metasurfaces under excitation by a 250 fs laser pulse resonant with the molecular $0-0$ transition, $\Omega_{\text{mol}} = 1.85$ eV, with a peak amplitude of $0.07$ V/nm. The vertical magenta dashed lines in all panels indicate the temporal end of the incident pulse. (a) Temporal evolution of the excited electronic state population for different configurations: TH (black line), SRR (red line), CSRR (blue line), one-dimensional cavity setup from Fig. \ref{fig:1}c (green line), and molecules in free space (purple line). (b) Time-dependent expectation value of the rotational quantum number $\langle J \rangle$. (c) Temporal evolution of molecular alignment relative to the $z$ axis, quantified by $\langle \cos^2(\theta) \rangle$. (d) Temporal evolution of molecular alignment relative to the $x$ axis, quantified by $\langle \cos^2(\phi) \rangle$.}
		\label{fig:9}
	\end{figure}
	
	Figure \ref{fig:9} analyzes the electronic relaxation dynamics of molecules following strong pump excitation at frequency $\Omega_{\text{mol}}$ with the incident polarization along $x$-axis. The molecular dynamics at the three plasmonic interfaces exhibits significantly lower relaxation rates compared to the dynamics in a one-dimensional resonant Fabry-Perot cavity under identical conditions. Multiple relaxation regimes emerge, characterized by an initial fast exponential decay transitioning to a notably slower decay at 750 fs. The molecules on the TH metasurface achieve the highest electronic excitation of 0.3 at 180 fs.
	
	The rotational dynamics, presented in Figs. \ref{fig:9}b, \ref{fig:9}c, and \ref{fig:9}d, reveals distinct behavior. The ensemble-averaged rotational quantum number $\langle J \rangle$ for molecules on metasurfaces continues to evolve temporally after the pump pulse, indicating persistent orientational dynamics. The alignment relative to the $z$-axis, quantified by $\langle \cos^2(\theta) \rangle$, follows a similar temporal evolution, with molecules exhibiting significant alignment at later times compared to the Fabry-Perot cavity configuration.
	
\begin{figure}
		\centering
		\includegraphics[width=0.9\linewidth]{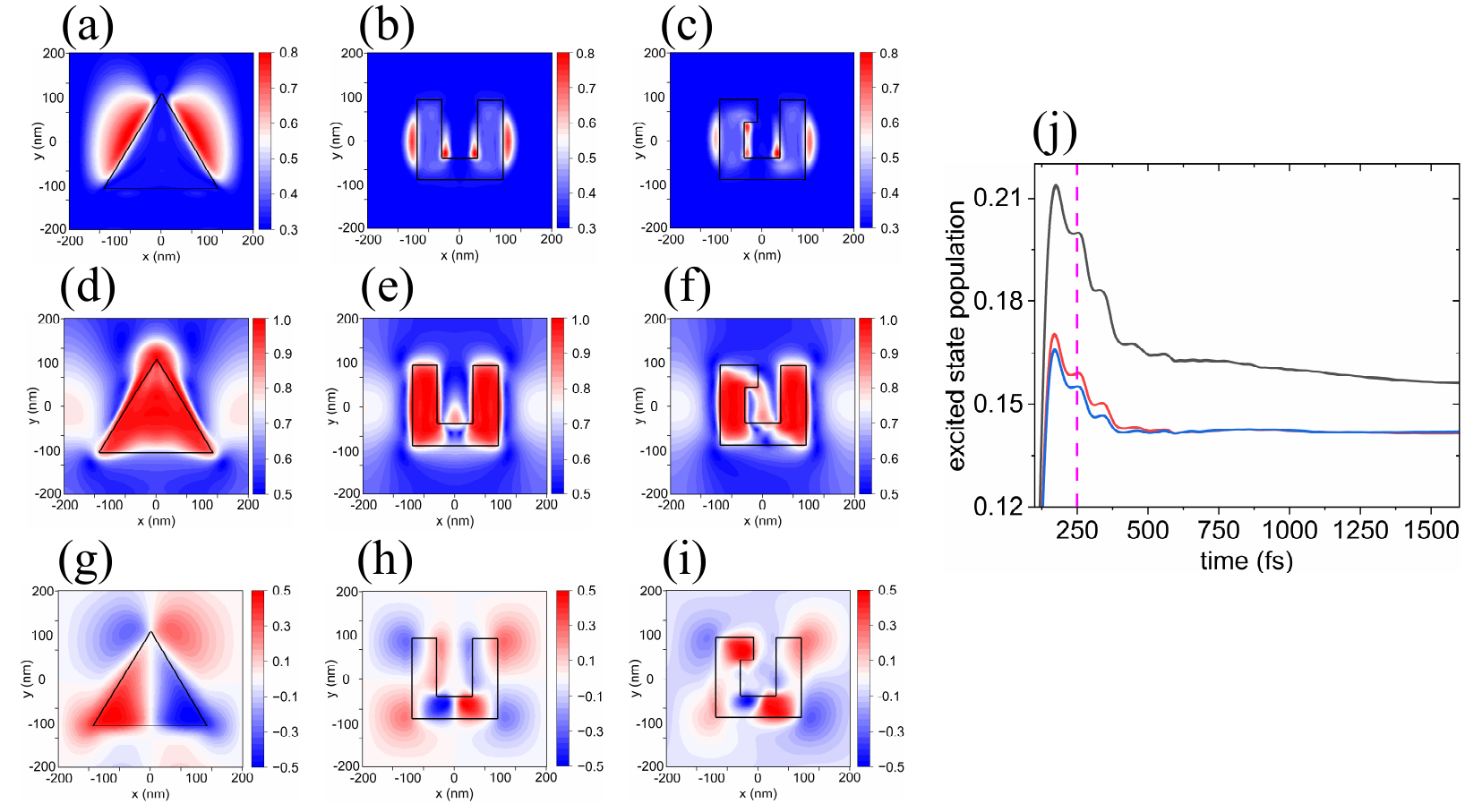}
		\caption{Three-dimensional simulations of molecules coupled to three metasurfaces under excitation at 1.89 eV. Spatial distributions at a detection plane 16 nm above the metal surface bisecting the molecular layer and recorded at 180 fs (maximum of the electronic excitaiton): $\langle \cos^2(\theta) \rangle$ (upper row, (a)-(c)), $\langle \cos^2(\phi) \rangle$ (middle row, (d)-(f)), and $\langle M \rangle$ (lower row, (g)-(i)). (j) Temporal evolution of excited electronic state population for molecules pumped at 1.89 eV for TH array (black line), SRR configuration (red line), and CSRR array (blue line). The vertical magenta dashed line indicates the end of the pump pulse. }
		\label{fig:10}
	\end{figure}

	The three-dimensional geometry introduces an additional alignment parameter $\langle \cos^2(\phi) \rangle$, where $\phi$ denotes the azimuthal angle between the $x$-axis and the molecular axis in the laboratory frame defined in Fig. \ref{fig:4}b. For initial conditions corresponding to $J=0$ and $M=0$, $\langle \cos^2(\phi) \rangle=\frac{1}{2}$. The temporal evolution shown in Figs. \ref{fig:9}c and \ref{fig:9}d demonstrates complex three-dimensional rotational dynamics.  Interestingly, the orientation in $xy$ plane is briefly stopped at the end of the pump, but then continues at later times following a similar trend observed in $\langle \cos^2(\theta) \rangle$. The highly spatially inhomogeneous EM field intensity and polarization distribution induce position-dependent ro-vibrational dynamics across the molecular ensemble, which is scrutinized in Fig. \ref{fig:10}.
	
The intriguing alignment dynamics observed long after the initial pump pulse stem from persistent EM fields in the system, which continue to influence molecular excitation and orientation states. Fig. \ref{fig:9}a reveals a striking difference in electronic relaxation rates: molecules on metasurfaces exhibit significantly slower relaxation compared to their free-space counterparts. This prolonged electronic excitation manifests in the rotational dynamics, as demonstrated in Figs. \ref{fig:9}b-d, where the rotational pumping process unfolds at a markedly reduced rate. This temporal modification of both electronic and rotational dynamics highlights the profound influence of metasurface-induced EM environments on molecular quantum states.
	
	Figures \ref{fig:10}a-i examine the spatial dependence of molecular orientation and the projection of angular momentum on the $z$-axis, $\langle M \rangle$, at the time of maximum electronic excitation (180 fs). The system is excited by a pulse centered at 1.89 eV, corresponding to the second sharp peak beyond the $0-0$ transition frequency in Fig. \ref{fig:7} (analogous to $\Omega_3$ in Fig. \ref{fig:1}d). 
	
	The spatial distribution of molecular alignment reveals distinct patterns: molecules orient predominantly vertical to the metasurface's plane near the nanohole edges (panels (a)-(c)), while the in-plane orientation, shown in panels (d)-(f), is localized in regions spanning the holes. The spatial distributions of $\langle M \rangle$ (panels (g)-(i)) exhibit remarkable correlations with the EM field chirality density maps presented in Fig. \ref{fig:5}. This correlation manifests most clearly in the TH configuration, where two prominent regions of alternating $\langle M \rangle$ near the upper corner of the triangular hole correspond directly to regions of left/right chiral fields. While the bottom region of the hole displays stronger $\langle M \rangle$ contrast than observed in the chirality density, the sign distribution of $\langle M \rangle$ maintains correspondence with the chirality density $C$ shown in Fig. \ref{fig:5}. Similar correlations between $\langle M \rangle$ and $C$ are evident in the SRR and CSRR configurations. This behavior is due to the angular momentum conservation and selection rules in light-matter interactions. For example, left-circularly polarized light carries angular momentum of $+\hbar$ per photon. When such a photon is absorbed by a molecule, the change in the magnetic quantum number $M$ must follow the selection rule $\Delta M=+1$ to conserve angular momentum. Therefore, only transitions that increase $M$ by $+1$ are allowed, which explains why spatial spots with positive/negative chirality preferentially excite positive/negative $M$ states.
	
	The temporal evolution of electronic excitation for molecules pumped at 1.89 eV, corresponding to a sharp absorption peak (Fig. \ref{fig:7}), demonstrates relaxation dynamics analogous to those observed in Fabry-Perot cavities (Fig. \ref{fig:3}), as illustrated in Fig. \ref{fig:10}j. All three metasurfaces configurations exhibit multiple stabilization plateaus as the molecular ensemble dephases relative to the respective SPP modes. Early-time dynamics show revivals of macroscopic polarization, manifesting as rapid decreases in electronic excitation. The stabilization plateaus become less pronounced after 600 fs, indicating transition to a different relaxation regime. The stabilization mechanism stems from the intricate interplay between collective molecular polarization and cavity modes. The excited molecular ensemble undergoes periodic polarization revivals, alternating between configurations that are in-phase and out-of-phase with the resonant cavity mode. During out-of-phase periods, the molecular system becomes effectively detuned from the cavity resonance, minimizing energy exchange and creating temporary stabilization. Conversely, in-phase alignment facilitates enhanced coupling to the electromagnetic mode, leading to rapid relaxation phases. This effect is expected to be more pronounced in homogeneous systems characterized by uniform molecular density distributions and a well-defined resonant electromagnetic mode. A direct comparison between our Fabry-Perot cavity simulations and three-dimensional metasurface results supports this conclusion: the metasurface geometry exhibits significantly shorter stabilization plateaus compared to those observed in the Fabry-Perot configuration, where spatial homogeneity is better preserved.
	
	\section{Conclusion} \label{conclusion}
	Our self-consistent quantum model provides fundamental insights into the complex dynamics of strongly coupled molecule-cavity systems. The observed modifications in molecular relaxation pathways, particularly the persistent alignment and collective dephasing effects, demonstrate the profound influence of cavity coupling on molecular quantum states. The cavity-induced rotational pumping mechanism represents a novel approach for controlling molecular orientation, with potential applications in ultrafast molecular switches and quantum information processing. The discovery of relaxation stabilization through collective dephasing suggests new strategies for preserving quantum coherence in molecular systems. Furthermore, the predicted alterations in circular dichroism spectra of chiral metasurfaces under strong coupling open avenues for engineering enhanced light-matter interactions in nanophotonic devices and chiral sensing applications. These findings advance our understanding of molecular polaritonics and provide a theoretical framework for designing next-generation quantum optical devices and molecular control techniques. Future work should focus on experimental validation of these predictions and exploration of their technological applications in quantum technologies and molecular photochemistry. 
	\begin{acknowledgement}
		
		M.S. research is supported by the Air Force Office of Scientific Research under Grant No. FA9550-22-1-0175 and by the Office of Naval Research, Grant No. N000142512090. J.E.S. was supported by the National Science Foundation under Grant No. CHE-2422858. The research of A.N. is supported by the Air Force Office of Scientific Research under Grant No. FA9550-23-1-0368 and the University of Pennsylvania. Simulations were made possible through the DOD High Performance Computing Modernization Program and were performed on DOD HPC systems: Narwhal, Nautilus, and Raider.
		
	\end{acknowledgement}

	\bibliography{acs-achemso}
	
\end{document}